\documentstyle[12pt,aaspp]{article} \input{amssym.def} %
\input{amssym.tex} 
mydef.tex 
\def\accepted#1{\gdef\@accptdate{#1}} \accepted{\relax}
\def\journalid#1#2{\gdef\@jourvol{#1}\gdef\@jourdate{#2}}
\def\articleid#1#2{\gdef\@startpage{#1}\gdef\@finishpage{#2}}
\begin{document} 
\today} \title{The Vertical Distribution and Kinematics of HI and Mass
Models of the Galactic Disk}

\author{Sangeeta Malhotra}
\affil{Princeton University Observatory, Peyton Hall, Princeton, NJ
08544 \\ san@astro.princeton.edu }

\begin{abstract}
We present full modelling of tangent point emission of HI as seen in
the 21cm transition in the inner Galaxy.  We measure the scale height
and velocity dispersion of HI as a function of Galactic radius besides
the rotation curve and the z-centroid of the atomic layer.  The model
used takes into account emission from a large path length along the
line of sight, corresponding to an interval ($\Delta R$) of typically
$\le 1 \kpc$ in galactic radius; and is parametrized by the scale
height of the gas, the centroid in z, the rotation velocity and the
velocity dispersion; these parameters are assumed to be constant over
the interval $\Delta R$.  This model is then fit to the 21 cm surveys
of Weaver \& Williams(1974), Bania \& Lockman (1984), and Kerr et al.
(1986) to determine the best fit parameters.

The terminal velocity values are found to be in good agreement with
previous measurements.  The velocity dispersion is very constant with
radius at the theoretically expected value of 9-10 $\kms$.  The
Gaussian scale height of the HI layer increases with Galactocentric
radius. The centroid of the layer deviates significantly from
z=0. Apart from small local wiggles, the four quantities measured:
terminal velocity $V_T$, velocity dispersion $\sigma_v$, scale height
$\sigma_z$ and the centroid $z_0$ show similar variations in the first
and the fourth quadrants.

On balancing the turbulent pressure support of the layer against the
disk gravitational potential, we confirm that additional support is
needed for the HI layer.  The radial profile of the reduced midplane
mass density is an exponential with a scale length of $3.3 \pm 0.3
\kpc$.  This picture is consistent with a constant mass-to-light ratio
of the disk, and extra support for the HI layer which is constant with
radius in the inner Galaxy.

\end{abstract}

\section {Introduction}

The analysis of vertical distribution and kinematics of disks of
galaxies can tell us about the local mass density in the disk,
distinct from the integrated mass inside a radius, derived from a
rotation curve (Oort 1932). HI is a particularly good tracer in our
own Galaxy as well as other systems because of its ubiquity (giving a
radial mass distribution), ease of observation, and near isothermal
nature, and has been used to derive disk mass densities in our own as
well as other galaxies (Rupen 1990, Knapp 1987, Merrifield 1992). In
the simplest case one considers the turbulent pressure gradient of the
gas balancing the gravitational force in the z-direction. But the
atomic gas may be subject to other pressures, for example, magnetic,
cosmic ray or radiation pressure. The relative contributions of these
are essentially unconstrained. The only place where the mass densities
obtained from the analysis of HI vertical equilibrium can be verified
with a similar analysis is the solar neighborhood, where the vertical
distribution and kinematics of stars can give an independent measure
of the midplane mass density ($\rho_0$) as well as surface density
($\Sigma$) of the disk.

As a starting point for this study we measure {\it both} the scale
height and the velocity dispersion of HI as a function of Galactic
radius between the radii of $\simeq 3-8 \kpc$. The interpretation of
kinematics of gas in the Galaxy is complicated due to our embedded
perspective.  The ubiquity of 21 cm emission of HI, and severe
blending in the low Galactic latitudes add to the difficulty (see
Burton 1988, 1992 for reviews). On the other hand our Galaxy is the
only three dimensional system we can study, and study at a high
resolution as well.

For some purposes simplifying assumptions can be made for certain
lines of sight, for example, Galactic rotation is expected to have
negligible contribution to the velocity spread of the gas at high
latitudes and towards galactic center and anti-center.  Velocity
dispersion of HI has been measured at these locations (Radhakrishnan
\& Sarma 1980, Kulkarni \& Fich 1985, Lockman
\& Gehman 1991). In the inner galaxy, the tangent points too lend
themselves to analysis with fewer assumptions. In the first and the
fourth quadrant emission at extreme velocity comes from an annulus
that is tangential to the line of sight, so we know the distance to
the tangent points and can derive the rotation curve of the Galaxy
after correcting for the velocity dispersion of the gas (Kwee et
al. 1954,, Schmidt 1957). Knowing the distance we can also convert the
angular extent of the gas to scale height.

In this paper we carry out full modeling of the tangent point
emission to derive the velocity dispersion at the tangent points. The
problem of velocity crowding is dealt with by considering emission
from a long path length (about 1 $\kpc$) upto the tangent points.  We
model the tangent points in two dimensions, latitude and velocity.
The apparent latitude extent of gas gives an extra handle on the
relative distance of any parcel of gas, assuming that the scale height
varies slowly with Galactic radius.  The model is parametrized by the
rotational velocity, the velocity dispersion, the scale height and the
z centroid of the gas layer.  These parameters are measured
independently for the tangent point gas at each longitude, allowing us
to study their variation with Galactic radius. Similar modeling has
been carried out for molecular gas (Malhotra 1994, hereafter paper I).
The modeling procedure is described in section 2.1.

This model is then fit to the HI surveys of Weaver \& Williams (1973,
hereafter WW), Bania \& Lockman (1984, hereafter BL), and Kerr et
al. (1986, hereafter Parkes). The specifications and parameters of the
surveys are briefly enumerated in section 2.2

We try to estimate uncertainties on the derived parameters in section
2.4.  The errors are not Gaussian, and are dominated by asymmetries
and other systematic deviations from the simple model used here, with
the atomic gas layer represented by a single Gaussian z-profile and a
single velocity dispersion component. These uncertainties (and
possible systematics) are estimated by changing the velocity and
latitude ranges over which the fits are made.

The modeling of the data at every $1 \deg $ in longitude yields the
rotation curve, the scale height and the velocity dispersion of HI as
function of Galactic radius between 3-8 kpc, for both the first and
the fourth quadrant. Section 3 describes the resulting best fit
estimates, and some possible implications for Galactic structure. One
of the objectives of this study is to examine the vertical equilibrium
of the gas. The parameters derived here are used to do so using the
the thin-HI-layer approximation in section 4.1. In section 4.2 we
consider a more realistic disk potential and calculate the expected
z-profile for isothermal gas. We also estimate by how much the
thin-HI-layer approximation underestimates the midplane mass density.

\section {Method}

\subsection{Tangent points}

Assuming a circularly symmetric model of the galaxy, we can obtain
distances to the emission at extreme positive velocities in the first
quadrant and extreme negative velocities in the fourth quadrant. At
each galactic longitude $l$, extreme velocity emission comes from the
tangent point at a galactocentric radius of $R=R_0 |\sin l|$ (Figure
1) and a distance $d=R_0 \cos l$ from the sun. From the observed $b$
extent and the terminal (extreme) velocities one could calculate the
scale height ($z= R_0 \cos l \tan b$) and the rotation velocity. The
velocity profiles do not have a sharp cutoff due to the velocity
dispersion of the gas. Ideally the emission from the tangent point
T(b,v) is a bivariate Gaussian in altitude and velocity; but emission
from nearby radii is not well separated in velocity because of the
velocity dispersion of the gas.

The finite velocity dispersion has two main effects that the present
analysis takes into account. First it makes the velocity-to-distance
conversion fuzzy, so it is difficult to separate emission from $R$ (at
the tangent point) and that from $R^\prime (R^\prime=R+\Delta R)$,
where $\Delta R$ depends on the velocity dispersion. For cold gas
(zero velocity dispersion) a difference in velocity $\Delta V_T=7
\kms$ would correspond to $\Delta R \simeq 270/\sin(l) \pc$ near the
tangent point. While the radial interval is small the line-of sight
distance changes substantially, leading to velocity crowding (Burton
1971), so that T(b,v) is no longer a simple bivariate Gaussian in $v$
and $b$.

Celnik et al.(1979) derived the expression for the expected line shape
near terminal velocity taking into account emission from gas all along
the line of sight, and used it to identify tangent point emission. In
this paper we calculate the 2-dimensional (in latitude and velocity)
contours of emission from near the tangent point, i.e.  between $R$
and $R+\Delta R$, assuming the scale height, rotation velocity
$\Theta$ and velocity dispersion to be constant over $\Delta R$. The
expected contours have a distinct shape at the tangent point (Figure
2). The HI emission does not peak at the terminal velocity (line of
sight velocity for the tangent point gas), but at a lower velocity.

Consider emission seen at velocities $V$ close to the terminal
velocity $V_T$; ($\Delta V = |V_T-V|\ll \Theta$). An observer's line
of sight intersects the annulus at radius $R^{\prime}$ at two points
(Figure 1). The distances to the points of intersection, $r_1$ and
$r_2$ are given by
\begin{eqnarray}
r_{\scriptstyle 1 \atop \scriptstyle 2}=R_0 \cos l\mp
R^{\prime}\sqrt{1-\frac{\sin^2 l}{(R^{\prime}/R_0)^2}} \nonumber \\ & & \\
\Rightarrow r_{\scriptstyle 1 \atop \scriptstyle 2}
=R_0 \cos l \mp R_0 |\sin l| \sqrt{\frac{\Theta ^2}{(\Theta -\Delta V)^2}- 1}
\end{eqnarray}

For the same scale height, gas at the subtangent points will have
smaller and larger latitude extent than the tangent point gas,
corresponding to the far and the near part of the annulus at
$R^{\prime}$ $$\sigma_{z \scriptstyle 1 \atop \scriptstyle 2}(V)=\frac
{\cos l}{\cos l\,\mp \,|\sin l|\sqrt{2 \Delta V/\Theta}} \sigma_z(V_T)$$

Taking into account velocity crowding effects, the optical depth per
velocity interval is determined by the line of sight distance per unit
velocity interval.

\begin{equation}
\frac{dr}{dV}=\frac{R_0 \Theta^2 \sin l}{{(\Theta_0 \sin l+ V(R ^{\prime}))}^3}
{\left ({ \frac{\Theta^2}{(\Theta_0 \sin l+V(R^{\prime}))^2}-1}\right)}^{-1/2}
\end{equation}

Thus the emission in the velocity interval $\Delta V$ (corresponding to
$\Delta R = \Theta_0/R_0 |\sin l|$) near the terminal
velocity, keeping parameters, $\sigma_v$, $\sigma_z$ fixed over the
velocity interval is

\begin{equation}
T(b,v)=A\sum_{1,2}\int{\frac{1}{2 \pi \sigma_V \sigma_z(V)}
\exp{\left(-\frac{(v/\cos b-V)^2}{2 \sigma_V^2}-\frac{(R_0 \cos l \tan
b-z_0(V))^2}{2 \sigma_z(V)^2} \right)}\frac{dr}{dV} dV}.
\end{equation}

\subsection{Data}

The modelling is done for three Galactic plane HI surveys. The HI
survey by Weaver and Williams (1973, hereafter WW) covers latitudes
$b=-10 \deg {\rm \ to \ } b=10 \deg$, sampling every $\Delta b=0\deg
.25$ with a beamwidth of $36\arcmin$. The results from modelling this
survey are compared to the the results from Arecibo HI survey (Bania
\& Lockman 1984, hereafter BL), done with a beamsize of $4\arcmin$,
and latitude coverage $\pm 4 \deg$.  In the fourth quadrant we use the
Parkes survey (Kerr et al. 1986).  The beamwidth is $48 \arcmin$ and
the latitude coverage $\pm 10 \deg$.  The resolution of these surveys
is not insubstantial compared to the scale height of the gas. A
half-power-beamwidth of $18\arcmin$ corresponds to linear scale of 40
pc at the tangent point at the longitude $20 \deg$ and to 21 pc at
longitude $60 \deg$. The Arecibo survey offers higher resolution but,
smaller latitude, and longitude coverage ($ |b| < 4 \deg, 32 \deg < l
< 64 \deg$). We will see that for the midplane position it will useful
to compare the results from both the surveys.

The observed brightness temperatures are converted to HI column
densities assuming a constant spin temperature $T_{spin}=127 \k$.  We
derive parameters for the longitude range $20 \deg < |l| < 62
\deg$. The lower limits is to side-step the very noncircular
kinematics of the bar, and the upper one because this method becomes
unreliable near the solar circle.

\subsection{Fitting}

The model profile is calculated (Eqn. 4) as a function of latitude $b$
and velocity $v$ for each longitude and the best fit is determined by
minimizing absolute differences between the model and the data. This
perhaps is a more robust method than minimizing least-squares, given
the higher-than-Gaussian tails expected in the z distribution, and the
expected presence of HI clouds, self-absorption, absorption and
asymmetries, all of which contribute to non-Gaussian noise. The
minimization is done using the downhill simplex routine, `amoeba'
(Press et al. 1993). In almost all the cases we are able to find
reasonable fits to the data at the tangent points. Figure 2 shows some
examples of good, reasonable and bad fits from the three surveys.
Best fits models for all longitudes are published elsewhere (Malhotra
1994b).

The velocity range over which the fit is made is determined by the
width of the extreme velocity feature. For each line of sight the
spectra at all latitudes are summed to form a composite spectrum. The
peak at the highest velocity is identified as the terminal velocity
feature. The lower velocity at which the emission drops to 80\% of the
peak value is defined as $V_{\rm crit}$. We fit the emission seen at
velocities greater than $V_{\rm crit}$. The HI because of its high
velocity dispersion and ubiquity shows severe blending and it was not
possible to isolate the tangent point emission in a large fraction of
the cases. We identified the extreme (positive in the first quadrant
and negative in the fourth) velocity at which the emission dropped to
3 times the noise-level as $V_{\rm ext}$ and then $|V_{\rm
crit}-V_{\rm ext}|=(14+26 |\sin l|)$, the second term being used so
that the velocity intervals correspond to the same interval in
Galactocentric radius ($\simeq 1\kpc$).

To see how sensitive the parameters obtained are to a change in the
velocity range over which the fitting is done, we do the fitting for
different values of $V_{\rm crit}$ for the longitude $l=42\deg$. The
results of this test are described in the next section, along with
other tests for systematic errors.

\subsection {Error Analysis}

The assumptions made in the modeling of the tangent point gas are as
follows.  We have assumed that the four parameters: the scale height,
the velocity dispersion, the rotational velocity and the position of
the centroid of the gas with respect to the z=0 plane, remain constant
over the radial range we fit over, which is no more than $1 \kpc$. So
the parameters we obtain are averages over such a range in
Galactocentric radius. While deriving the models we have also
implicitly assumed that the midplane volume density $n_{HI}$ is not
varying systematically over that region. Also, the spin temperature of
the gas is assumed to be constant. We also assume circular motions,
apart from velocity dispersions. But we know that $n_{HI}$ varies on
all scales, and expect that $n_{HI}$ and velocity fields undergo large
scale variations at the spiral arms.

An additional source of error is that the beam sizes for the WW and
Parkes surveys are greater than $0.\deg 5 $, typical longitude
intervals over which we make `independent' estimates and see the
parameters change is $1 \deg$. What that means is that we are
averaging over the azimuthal (longitudinal) extent of the beam, as
well the line-of-sight interval. While the latter is accounted for in
the model (by changing the b-extent corresponding to the same scale
height, for example), the former is not accounted for. Notice that the
model-fits are substantially better for the BL survey (Figure 2),
where the resolution is better. The models however describe the gross
features of the gas distribution fairly well, and the parameters
derived from the BL survey agree with those from WW survey (Figures
5, 6, 7, 8).

We find that the chi-square per degree of freedom: ${\chi^2}/{N}$, for
the fits is almost always greater than 1. Moreover, looking at the
residuals left after subtracting the model from the data, we find that
the quality of fit is always dominated by systematic errors, for
example asymmetries in the b-extent about the centroid,
high-than-Gaussian tails at high $z$ etc.

To test for systematic sources of error, we do the model fitting for
different ranges in b, for two lines of sight $l=35\deg$ and
$l=25\deg$. Figure 3 shows the best-fit parameters for the different
z-ranges, the dispersion in these values gives some estimate of the
error-bars on the parameters. The velocity dispersion is found to be
uncertain by $1 \kms$, the terminal velocity by $2 \kms$, the
midplane position by $ 2-8 \pc$, and the scale height is uncertain by
$15-20 \pc$.

The other noticeable effect is the variation of the parameters between
$b>0$ and $b<0$. The scale height is larger on one side of the plane
than the other. There is also no significant variation of velocity
dispersion with distance from the plane. That the HI velocity
dispersion does not change with height above the plane is a strong
indication that the gas is very nearly isothermal.

We also test for the dependence of the best fit parameters on the
velocity range over which the fitting is done. Figure 4(b) shows the
best fit parameters plotted against the $V_{crit}$ where the fitting
starts. The variations in parameters are comparable to the
uncertainties quoted in the last paragraph. Except the scale height
$\sigma_z$, which stays constant for small changes ($\Delta V_{crit} <
10 \kms$) in the $V_{crit}$ and shows a systematic rise as we decrease
the $V_{crit}$ to include emission from a larger part on the
line-of-sight gas (Figure 4(a)). This is reasonable as we are
including more gas along the line of sight in the average, and this
gas being at a larger Galactocentric radius has a greater scale
height, since the scale height increases with Galactocentric radius.

Going through rough calculations it is easy to see that the velocity
interval between the terminal velocity of $71 \kms$ and the most
extreme $V_{crit}=34 \kms$ corresponds to $\Delta R \simeq 2
\kpc$. As we shall see in the next section (describing the results),
the scale height increases by $\simeq 40 \pc$ in that radial interval,
and we see that the estimate of the average scale height increases by
$\simeq 20 \pc$ when we include gas from that annulus. So what if
$V_{crit}$ is too low and includes too large a range in $\Delta R$?
For small errors ($< 10 \kms$) the results are not very erroneous, and
the test case in Figure 4(b) is an extreme example of low
$V_{crit}$. If a large systematic mistake is made in a large fraction
of longitudes we may be underestimating the increase of $\sigma_z$
with Galactic radius.

\section {Results}
\subsection {Midplane $Z_0$}

The position of the centroid of the HI layer deviates significantly
from $Z=0$, even in the inner Galaxy. Lockman (1977) has pointed out
the large scale Z-displacements of population I objects in the inner
Galaxy from the Z=0 plane. For the inner Galaxy, the Z displacement
can most reliably be measured at the tangent points. At other places
we see a superposition of the near and far layers. The midplane
positions $Z_0$ of the gas layer at the tangent points in the first
and the fourth quadrants are shown in Figure 5. Also shown superposed
are the midplane deviations of the molecular gas as measured by CO
emission (Malhotra 1994, Bronfman et al. 1988).

The deviations of the molecular gas layers are easily detected because
of the smaller scale height of the layer ($\simeq 50 \pc$ compared to
$> 100 \pc$ for HI scale height), and because of the high resolution
of the surveys ($100 \arcsec$ - $7.5 \arcmin$). The WW and Parkes
surveys have linear resolutions of several tens of parsecs depending
on the longitude of the tangent point. The $Z_0$ values determined
from BL survey with 9 times smaller beamsize, agree well with those
determined from WW survey (Figure 5). This leads us to have confidence
in $Z_0$ measured in the fourth quadrant.

The midplane positions of the atomic and molecular gas show similar
deviations from the $Z=0$ plane. In the first quadrant the maximum
deviation in $Z_0\simeq -50 \pc$, occurring at about $42 \deg $
longitude.  Moreover $Z_0$ seems to be mostly negative, changing sign
at $l=53 \deg$. In the fourth quadrant the opposite is true. Most of
the gas layer has a positive deviation from $Z=0$, changing sign at
$l= -55 \deg$. Maximum excursion is $Z_0 \simeq 50 \pc$ at $l=33 \deg$.
Recall that the outer galaxy warps to positive $Z$ in the first
quadrant, and negative $Z$ in the fourth. Thus the inner Galaxy
displacements though complicated are largely opposite to the outer
Galaxy warp.

\subsection {Scale height}

The scale height is estimated by fitting a model which assumes a
Gaussian profile for the vertical distribution of HI, with the
midplane position $Z_0$ and scale height as parameters. There are many
reasons and observations to believe that the gas layer is not
adequately described by a single Gaussian. An isothermal population
tracing an external potential is expected to have a Gaussian profile
near the plane. There are observations to indicate that there is more
HI at higher Z ($> 500 \pc$) than predicted by a single Gaussian fit
made near the plane (Lockman 1984). However it is not possible to
reliably estimate all the parameters required for a more complicated
model, say with two scale heights and as many values of velocity
dispersions. A single Gaussian profile in $z$ describes about 90\% of
the HI in the inner Galaxy (Lockman 1984).

As mentioned before the beamwidths in the Parkes and WW surveys are
non-negligible compared to the scale height. Moreover the beamwidths
correspond to larger physical sizes for tangent points at greater
distances from the sun (at small Galactic radii). The best-fit scale
heights are corrected for this effect by subtracting the beam
half-widths in quadrature. As seen in Figure 6, the scale height
increases with Galactic radius, for $R > 0.5 R_0$, in both the first
and the fourth quadrant. The variation of the scale height with $R$ is
similar in the two quadrants. This result is only marginally
inconsistent with the constancy of scale height with Galactic radius
(Lockman 1984) and a linear rise of $\sigma_Z$ with Galactic radius
(Knapp 1987). HI in external edge-on galaxies also show increase in
scale-height with radius (Rupen 1990).

In the first quadrant the BL and the WW data show the same scale
heights (after correction for beamsize) for longitudes $l \le 48.\deg6
$. For higher longitudes the BL survey shows a smaller scale height
than the WW data. This is due to the limited coverage of the BL
survey, combined with the increasing apparent scale height at higher
longitudes as the tangent points get closer to the position of the
sun. Fitting to WW data restricted to latitude extent of the BL survey
gives similar values as a fit to the BL data.

\subsection {Terminal velocity}

In Figure 7 we show the best fit terminal velocities for the first and
the fourth quadrants. The curve for the first quadrant is compared to
a similar curve derived by Gunn, Knapp \& Tremaine 1979, from the 1 K
contour at the extreme velocities, after correcting for a velocity
dispersion of 9 kms.  The two curve agree fairly well, reassuring us
that the present tangent point modelling does not grossly
over/underestimate the terminal velocities.

There exists asymmetry between the first and the fourth quadrant
terminal velocity curves. This difference is not easy to understand as
a simple outward motion or as other systematic (Blitz \& Spergel 1991,
Kerr 1962)

\subsection {Terminal velocity}

In Figure 7 we show the best fit terminal velocities for the first and
the fourth quadrants. The curve for the first quadrant is compared to
a similar curve derived by Gunn, Knapp \& Tremaine 1979, from the 1 K
contour at the extreme velocities, after correcting for a velocity
dispersion of 9 kms.  The two curve agree fairly well, reassuring us
that the present tangent point modelling does not grossly
over/underestimate the terminal velocities.

There exists asymmetry between the first and the fourth quadrant
terminal velocity curves. This difference is not easy to understand as
a simple outward motion or as other systematic (Blitz \& Spergel 1991,
Kerr 1962)

\section {Vertical equilibrium of the HI layer}

Considering the condition for vertical equilibrium of the atomic gas
in the galactic disk, in the simplest case of a single isothermal gas,
the vertical component of the gravitational force is balanced by the
pressure due to turbulent motions of the gas and due to magnetic
fields etc.
\begin{equation}
K_z=-4 \pi G\Sigma(z) \rho_{gas}= \frac {d}{dz} (P_{kinematic}+P_{magnetic}+
P_{cosmic rays} +P_{radiation} \ldots)
\end{equation}

The scope of this paper is limited to analyzing the kinematics of the
HI layer and its vertical structure, so we will only consider
gravitational forces on the gas and the kinetic support.  Very near
the plane we can assume the total disk mass density to be nearly
constant $\rho_0$, and the surface density $\Sigma(z)=\rho_0
z$. Considering only kinematic pressure support the above equation
reduces to
\begin{equation}
 \frac{d}{dz}\rho_{gas} \sigma_v^2=-4\pi G\rho_0 z \rho_{gas}
\end{equation}
and the z-profile of the gas is a Gaussian $\rho_{gas}(z)\propto
exp(-z^2/2 \sigma_z^2)$ with the scale height
$\sigma_z^2=\sigma_v^2/4\pi G\rho_0$.

The total midplane total mass density $\rho_0=\sigma_v^2/4\pi
G\sigma_z^2$, and is determined for each longitude, from the
measurements of the scale height and the velocity dispersion, using
the thin-layer approximation. We have also assumed that the velocity
dispersion of HI is isotropic, so the azimuthal velocity dispersion we
measure is equal to the vertical velocity dispersion. This is
justified since HI is a diffuse collisional medium. Support for this
assumption also comes from the isotropy of the velocity dispersion of
young stars who reflect the kinematics of the parent interstellar gas
(cf Paper I). Figure 9 shows the radial profile of the midplane mass
density $\rho_0(R)$ as a function of radius R for both the first and
the fourth quadrants. An exponential profile (in radius) is fit to
$\rho_0$ measurements at each longitude and yields a scale length of
$3.4 \pm 0.3 \kpc$ in the first quadrant, and $3.1 \pm 0.3 \kpc$ in
the fourth.

The scale length of the midplane mass density is the same as one of
measurements of the scale length of the light in the exponential disk
(de Vaucouleurs \& Pence 1978), suggesting a constant mass to light
ratio at least in the inner Galaxy. There are various estimates of the
scale length of our galaxy ranging from $1.8-6 \kpc$(see Kent, Dame \&
Fazio 1991, for a list and discussion). Most estimates from infrared
studies give a scale length of the light volume density of about $3
\kpc$.

The uncertainty in the scale length of disk-mass-density is determined
by the bootstrap method. This method was used because it is difficult
to assign realistic error bars to each measurement of scale height
$\sigma_z$ and velocity dispersion $\sigma_v$ (cf. previous
section). Since $\rho_0$ depends on the square of $\sigma_v$ and
$\sigma_z$, the error bars on $\rho_0$ are even more
ill-determined. With more than 50 determinations of $\rho_0(R)$ in
each of the quadrants the bootstrap method easily yields the
uncertainty in the scale length.

Merrifield (1992) has estimated the mass density profile of the disk
to be an exponential with a scale length of about $5 \kpc$ from his
measurements of the scale height and assuming the velocity dispersion
of the HI is $10 \kms$. Using the scale height determination from the
last section with $\sigma_v=10 \kms$, gives a scale length of $4.6
\kpc$ in the first quadrant and $3.3 \kpc$ in the fourth. These values
differ from each other and from the scale lengths calculated above
using the measured $\sigma_v$ and $\sigma_z$ by more than the
error-bar on the scale length. We may interpret this discrepancy as
the true measure of the uncertainty in the scale length or as an
indication that the individual measurements of $\sigma_v$ and
$\sigma_z$ show local variations that are significant when estimating
mass density profile $\rho_0(R)$.

Extrapolating $\rho(0)$ to solar radius gives a a midplane mass
density of $0.03 \msun pc^{-3}$ at least a factor of $\simeq 3.5$
smaller than the midplane mass density inferred at the sun's position
from stellar kinematics, (which is also uncertain by a factor of two)
(Bahcall et al. 1992, Bahcall 1984a, Bahcall 1984b, Bahcall 1984c,
Bienaym\'e et al. 1987, Kuijken \& Gilmore 1989a, Kuijken \& Gilmore
1989b, Kuijken \& Gilmore 1989c, Kuijken \& Gilmore 1991a, Kuijken \&
Gilmore 1991b, Kuijken 1991). This implies that extra pressure is
needed to support the HI layers to the height we see them at.

The candidates for extra support are pressure gradients of one or more
of the following: pressure due to high velocity dispersion
component(s) of the gas, magnetic pressure, cosmic ray pressure, and
radiation pressure. These candidates are examined in detail by Boulares
\& Cox (1990) and Lockman \& Gehman (1991). Given that the scale length
of $\rho_0(R)$ is the same as the scale length of light in the disk,
constant mass-to-light ratio of the disk is favored. The velocity
dispersion of HI remains constant at $9\pm 1 \kms$, showing that the
decrease in $\rho_0(R)$ is reflected in the scale height of the gas,
which increases with radius. This implies that the extra support for
HI, whichever source it comes from, should stay constant with radius.

\subsection{Some systematics}

We expect the HI to be in a Gaussian layer only if the gas layer is
much thinner than the stellar scale height, so we may assume an
effective constant stellar density with z. This is the approximation
used in the previous section. Now we calculate the z-profile of the
atomic gas in a realistic disk potential given that it has a velocity
dispersion of $\sigma_v=10 \kms$. Forsaking the approximation that the
gas is in a layer much thinner than the stellar layer, the z-profile
of the gas is given by

\begin{eqnarray}
 \frac{d}{dz}\rho_{gas}(z) \sigma_v^2=-4\pi G \rho_{gas} \int_0^z
{\rho(z^{\prime}) d(z^{\prime})}\nonumber \\ & & \\
\frac{d^2}{dz^2}\log(\rho_{gas}(z))=\frac{-4\pi
G}{\sigma_v^2}(\rho_{gas}(z)+\rho_{*}(z)+\rho_{molecular\  gas}(z))
\end{eqnarray}

The stellar distribution $(\rho_{*}(z))$ is assumed to be exponential
in $z$ with a scale height of $320 \pc$ (Bahcall \& Soneira 1984), and
the molecular gas profile in z is assumed to be a Gaussian with a
scale height of 60 pc. The midplane mass densities of the stellar and
molecular component are taken to be $0.072 \msun pc^{-3}$ and $0.021
\msun pc^{-3}$ respectively. HI midplane mass density is assumed to be
$0.021 \msun pc^{-3}$. Figure 10 shows the z-profile for the gas in
this potential.

The expected z-profile is very nearly a Gaussian with a scale height
of $144 \pc$, but it has slightly high tails (about $1.4\%$ of the
total gas lies beyond $3-\sigma_z \pc$). Using this scale height and
velocity dispersion of $10 \kms$ gives a midplane mass density $0.9
\msun pc^{-3}$ as opposed to the input of $0.11 \msun pc^{-3}$. The
thin-layer approximation thus leads to an underestimate of the
midplane mass density by about $25\%$.

\section{Conclusions and Summary}
The main results of this study  are as follows:

(a) The velocity dispersion of HI is measured to be constant at 9
$\kms$ across the inner Galaxy (Galactic radius of 3-8 $\kpc$).

(b) The scale height of the gas increases with the Galactic radius. It
increases by nearly 100 pc going from 3-8 $\kpc$ in Galactic radius.

(c) The midplane mass density shows an exponential decline with a
scale length of $3.4 \pm 0.3 \kpc$ and $3.1 \pm 0.3 \kpc$ in the first
and the fourth quadrant respectively. Assuming that it is not a
coincidence that the scale length of light in the disk is the same,
this indicates a constant mass to light ratio of the disk.

(d) The midplane mass density when extrapolated to solar radius, is
smaller by a factor of 3.5 compared to the value derived by Kuijken \&
Gilmore 1989. This indicates that the HI layer needs extra support
(magnetic, or cosmic ray pressures, radiation pressure) to keep it up
at the heights it is seen.

(e) Since the exponential decline in the midplane mass density
$\rho_0(R)$ with R, happens with the kinetic pressure support remaining
constant, the extra support for the layer should also remain constant
in the Galactic radius range of 3-8 $\kpc$, for the M/L ratio to
remain constant for the disk.

\acknowledgements
This paper forms a part of the author's Ph.D. thesis at Princeton
University. I thank G. R. Knapp for advice at all stages of the thesis
and F J Lockman for providing the Arecibo data from the HI survey by
Bania \& Lockman (1984). I would like to thank F J Lockman, M Rupen,
and R H Lupton for discussions and R. H. Lupton for his versatile
graphics package `SM'. This work was supported by NSF grant
AST89-21700 to Princeton University.

\vfill
\pagebreak

\parindent=0pt
Figure 1: The geometry of the tangent point emission. S indicates the
position of the sun, R is the Galactic radius at the tangent point,
and ${\rm R}^{\prime}$ is a sub-tangent point. The models of the
tangent point emission take into account the emission from the annulus
between R and ${\rm R}^{\prime}$.

Figure 2: Latitude-velocity maps of the 21 cm emission at the
longitudes $l=22\deg, 37\deg, 55\deg$ (WW survey), $42.\deg 5$ (BL
survey) and 290\deg (Parkes survey). The best fit models to the
tangent point emission are shown superposed on the data.  Contour
levels are defined at 10, 20, 30, .. 90\% of the peak temperature in
the best-fit model. The model shows an abrupt cutoff at the velocity
$V_{\rm crit}$, because the fitting is done only for $V > V_{\rm
crit}$; $V_{\rm crit}$ being the velocity at which the tangent point
emission is 80\% of the peak value or chosen according to other
criterion described in Section 2.3.

Figure 3: The effect of changing the b-range over which the fitting is
done for the longitude $l=25\deg$ and $35 \deg$. The latitude range is
plotted on the y axis, and the vertical bars indicate the b-range over
which the fitting was done. The parameters (with the exception of
scale height $\sigma_z$) are fairly insensitive to the b-range of the
fit. The scale height is different on the north and the south side,
ranging from 120 pc in the north to 180 pc in the south. The small
midplane displacement to the north does not fully account for
this. This shows that the layer is fairly asymmetric.

Figure 4(a): Latitude-velocity maps of 21 cm emission of HI at the
longitude l=42$\deg$. The dotted vertical lines show the various
values of $V_{\rm crit}$. The fitting is done for emission at
velocities $V > V_{\rm crit}$. The solid line shows the tangent point
velocity, and is not coincident with the peak in emission (Section
2.1).

Figure 4(b): The best fit parameters of the model for the longitude
$l=42 \deg$: the terminal velocity $V_T$, the velocity dispersion
$\sigma_v$, the centroid in z - $Z_0$ and the scale height $\sigma_z$
are plotted as a function of the cutoff velocity $V_{\rm
crit}$. Emission at velocities $ > V_{\rm crit}$ was modeled. This
figure shows that the parameters of the model are not very sensitive
to small changes in the boundaries of the region they are fit. Only
the scale height shows a systematic increase. This is discussed in
section 2.4.

Figure 5: The z location of the centroid of the midplane in the first
(open circles: Weaver \& Williams survey, triangles: Bania \& Lockman
survey) and the fourth quadrant (open squares: Parkes survey by Kerr et
al.). The solid lines are the midplane positions of the molecular gas
layer as traced by CO emission (Malhotra 1994, Bronfman 1988). There
is a resonable agreement between the midplane deviations of the two
phases of cold neutral gas.

Figure 6: The scale heights of HI layer in the first and the fourth
quadrants (open circles: WW survey, triangles: BL survey, open
squares: Parkes survey). The HI shows similar behaviour in the first
and the fourth quadrants, flaring at $\simeq 0.6 R_0 (5 \kpc)$. The
scale heights obtained from tangent point modeling have been corrected
for broadening by the telescope beams (Section 3.2)

Figure 7: The terminal velocity $V_T$ for the tangent point gas at
different Galactic longitudes $l$ (hence Galactic radii $R/R_0=
\sin(l)$). The solid line shows the terminal velocity for HI from
Gunn et al. 1979 (open circles: WW survey, triangles: BL survey, open
squares: Parkes survey).

Figure 8: The velocity dispersion of HI as modelled from tangent
point emission (open circles: WW survey, triangles: BL survey, open
squares: Parkes survey). The velocity dispersion is seen to be fairly
constant at $\simeq 9 \kms$. The scatter in the individual values and
other estimates (see section 2.4) yield an uncertainty of $1 \kms$ in
this value.

Figure 9: The midplane mass density of the disk
$\rho_0=\sigma_v^2/4\pi G\sigma_z^2$ (open circles: WW survey,
triangles: BL survey, open squares: Parkes survey). Exponential disk
models are fitted to the WW data in the first quadrant and Parkes data
in the fourth. The best fit scale-lengths are $3.4 \pm 0.3 \kpc$ and
$3.1 \pm 0.3 \kpc$ for the first and the fourth quadrant
respectively. The error bars on the scale-length are calculated from
bootstrapping the individual measurements of $\rho_0(R)$.

Figure 10: The z-profile of isothermal HI with a velocity dispersion
of $\sigma_v=10 \kms$ is calculated numerically in a realistic
potential: Solid line (cf Section 4.1). This profile has
higher-than-Gaussian tails at high z- about 1.4\% of the gas lies
beyond $3-\sigma_z (\simeq 500 \pc)$. The closest Gaussian profile is
marked by the dashed line.

\end{document}